\documentclass[conference]{IEEEtran}
\IEEEoverridecommandlockouts
\usepackage{cite}
\usepackage{amsmath,amssymb,amsfonts}
\usepackage{algorithmic}
\usepackage{graphicx}
\usepackage{textcomp}
\usepackage{bmpsize}
\usepackage{lipsum}
\usepackage{array}
\usepackage{booktabs}
\usepackage{caption}
\usepackage{adjustbox}
\usepackage{amssymb}
\usepackage{amsmath}
\usepackage{fdsymbol}
\usepackage{enumerate}
\usepackage{xcolor}
\usepackage{subfig}
\usepackage{float}
\usepackage{balance}
\usepackage{soul}
\usepackage{multirow}
\usepackage{enumitem}
\usepackage{comment}
\usepackage{wasysym}

\usepackage[colorlinks=false,urlcolor=black, hidelinks]{hyperref}
\def\BibTeX{{\rm B\kern-.05em{\sc i\kern-.025em b}\kern-.08em
    T\kern-.1667em\lower.7ex\hbox{E}\kern-.125emX}}
\begin{document}

\newcommand{\Tdot}{$\medblackcircle$}
\newcommand{\TDot}{$\medcircle$}
\newcommand{\Toco}{$\oast$}
\newcommand{\Temp}{$\Circle$}
\newcommand{\Tsqu}{$\medblacksquare$}
\newcommand{\TSqu}{$\medsquare$}
\newcommand{\Tsla}{$\oslash$}
\newcommand{\Thdot}{$\LEFTcircle$}
\newcommand{\Twdot}{$\oast$}
\newcommand{\Tdia}{$\medblackdiamond$}
\newcommand{\TDia}{$\meddiamond$}
\newcommand{\Ttri}{$\medblacktriangleup$}
\newcommand{\TTri}{$\medtriangleup$}
\newcolumntype{R}[2]{%
    >{\adjustbox{angle=#1,lap=\width-(#2)}\bgroup}%
    l%
    <{\egroup}%
}
\newcommand*\rot{\multicolumn{1}{R{60}{1em}}}
\newcommand*\rota{\multicolumn{1}{R{90}{1em}}}
\newcommand{\etal}{\emph{et al.}\xspace}

\newcommand{\rev}[1]{\textcolor{teal}{#1}}
\newcommand{\federico}[1]{\textcolor{red}{Federico: #1}}
\newcommand{\giovanni}[1]{\textcolor{purple}{Giovanni: #1}}
\newcommand{\edit}[1]{\textcolor{red}{#1}}

\newcommand{\parag}[1]{\noindent\textbf{#1. }}

\setlist[itemize]{noitemsep, topsep=1pt}
\captionsetup{belowskip=2pt}

\title{Assessing the Use of Insecure ICS Protocols via IXP Network Traffic Analysis}

\author{
    \IEEEauthorblockN{Giovanni Barbieri\IEEEauthorrefmark{1}, Mauro Conti\IEEEauthorrefmark{1}, Nils Ole Tippenhauer\IEEEauthorrefmark{2}, Federico Turrin\IEEEauthorrefmark{1}}
    \IEEEauthorblockA{\IEEEauthorrefmark{1}University of Padua, Department of Matemathics\\
    Padua, Italy \\
    brbgiovanni@gmail.com, \{conti, turrin\}@math.unipd.it}
    \IEEEauthorblockA{\IEEEauthorrefmark{2}CISPA Helmholtz Center for Information Security\\
    Saarbrücken, Germany\\
    tippenhauer@cispa.de}
}

\maketitle

\begin{abstract}
Modern Industrial Control Systems (ICSs) allow remote communication through the Internet using industrial protocols that were not designed to work with external networks. To understand security issues related to this practice, prior work usually relies on active scans by researchers or services such as \emph{Shodan}. While such scans can identify publicly open ports, they cannot identify legitimate use of insecure industrial traffic. In particular, source-based filtering in Network Address Translation or Firewalls prevent detection by active scanning, but do not ensure that insecure communication is not manipulated in transit. 

In this work, we compare \emph{Shodan}-only analysis with large-scale traffic analysis at a local Internet Exchange Point (IXP), based on sFlow sampling.
This setup allows us to identify ICS endpoints actually exchanging industrial traffic over the Internet. Besides, we are able to detect scanning activities and what other type of traffic is exchanged by the systems (i.e., IT traffic).
We find that \emph{Shodan} only listed less than 2\% of hosts that we identified as exchanging industrial traffic, and only 7\% of hosts identified by \emph{Shodan} actually exchange industrial traffic. Therefore, \emph{Shodan} do not allow to understand the actual use of insecure industrial protocols on the Internet and the current security practices in ICS communications.
We show that 75.6\% of ICS hosts still rely on unencrypted communications without integrity protection, leaving those critical systems vulnerable to malicious attacks.
\end{abstract}


\section{Introduction}\label{sec:intro}
Industrial Control Systems control and monitor industrial and critical infrastructure such as power grids, nuclear and chemical plants, water treatment systems and buildings. According to the Purdue model~\cite{williams1994purdue}, which represents the reference architecture model for the ICSs, these systems were supposed to operate on air-gap environments. However due to the rise of the Internet and Ethernet technologies, the industrial business moved towards the connection to external networks, opening dangerous vulnerability surfaces. This digitization process requires the adaptation of legacy protocols that were not designed with security features (e.g., Modbus and DNP3) to the TCP/IP stack to allow operations such as remote communication. The lack of encryption and authentication can be exploited for a malicious purpose such as data exfiltration, impersonification, and service distruption.
The research interest and effort on ICS security increased with the number of incidents and the involvement of critical-infrastructures, from the most famous Stuxnet~\cite{stuxnet} that affected an Iranian nuclear power plant to the more recent Triton~\cite{triton} that targeted Schneider Electric products. 
As reported by~\cite{kaspersky2019}, the attacks over the ICSs are growing year after year, and the number of vulnerable devices is still very high. Many researchers estimated the vulnerable ICS landscape by looking for well-known ICS ports exposed to the Internet, by actively performing scanning of the IPv4 address space~\cite{Mirian2016,Feng2016} or leveraging public available information~\cite{Al-Alami2018,Twente2019} gathered by third-parties such as \emph{Shodan}~\cite{shodan} and \textit{Censys}~\cite{censys}. \emph{Shodan} and \textit{Censys} are cloud services which allow the user to scan, discover, and query the devices connected to the Internet.
However, many ICSs could not be indexed due to the presence of network devices like Network Address Translations (NATs) or Firewalls, leading to an incomplete estimation of the vulnerable ICSs. In~\cite{Nawrocki2019} the authors analyzed the unprotected industrial traffic transmitted over the Internet, gathering information about the security status of the host systems.

In this paper, we investigate the practical use of ICS protocols over the Internet, most importantly to learn about security issues. Moreover, we aim to determine if scanning activities (such as results from \emph{Shodan}) provides accurate estimates of the use of ICS protocols. To achieve this, we present a framework to identify the ICS host transmitting industrial traffic over the Internet, based on traffic observed at an IXP. We compare and correlate the obtained results with the information from a prominent scanner: \emph{Shodan}. We use \emph{Shodan} due to its popularity in related works~\cite{Al-Alami2018,Twente2019}, and because it supports significantly more ICS protocols with respect to other services (e.g., Censys).
Our analysis shows that ICS hosts identified by \emph{Shodan} are rarely actively using ICS protocols, and such measurements do not estimate the use of insecure protocols on the Internet. Instead, IXP (or ISP)-based measurements are required. Our results also provide insights into the volume and type of legitimate and unprotected ICS communication found on Internet links.

We summarize our contributions as follows:
\begin{itemize}
    \item We implement and validate a framework to identify legitimate industrial traffic and scanning activities based on sampled traffic passively gathered at an IXP. 
    \item We demonstrate that our method allows to obtain a detailed understanding of security practices in ICS, in particular related to site-to-site communication with industrial protocols.
    \item We show and confirm the current security issues, vulnerabilities, and exposure related to the Industrial Control Systems implementation.
\end{itemize}

\section{ICS Traffic Analysis Through IXP}\label{sec:experiments}

Our analysis relies on \emph{VSIX}~\cite{vsix}, a local IXP which manages the traffic circulating in the North East of Italy.
This environment allows to collect packets circulating through the Internet in a secure and privacy respectful way. In the following, Section~\ref{subsec:ASIXP} recalls the definition of Autonomous System (AS) and IXP which represent the main entities of our system model. We detail our system model in Section~\ref{subses:system_model}, while in Section~\ref{subses:sampling_model} we present the theory behind the packet sampling process. Then, Section~\ref{subses:questions} and Section~\ref{subses:pipeline} outline respectively, the research questions of this paper and the framework we implemented to address the proposed questions. 

\subsection{ASs and IXPs}\label{subsec:ASIXP}

An Autonomous System (AS) is a group of IP prefixes under the control of a single well-defined administrative authority that defines the routing policies, typically an Internet Service Provider (ISP), uniquely identified by its Autonomous System Number (ASN). The routing within an AS is allowed via Interior Gateway Protocol (IGP), while the communication with other ASs relies on the Border Gateway Protocol (BGP). An IXP is a network facility that enables the interconnection and exchange of Internet traffic between more than two independent ASs according to their BGP routing configurations. Its typical architecture consists of a single or multiple switches connected to the border routers of the adherent ASs, ensuring benefits in terms of bandwidth, costs, and latency.

\subsection{System Model}\label{subses:system_model}

\begin{figure}[t]
\centering
\includegraphics[width=\columnwidth]{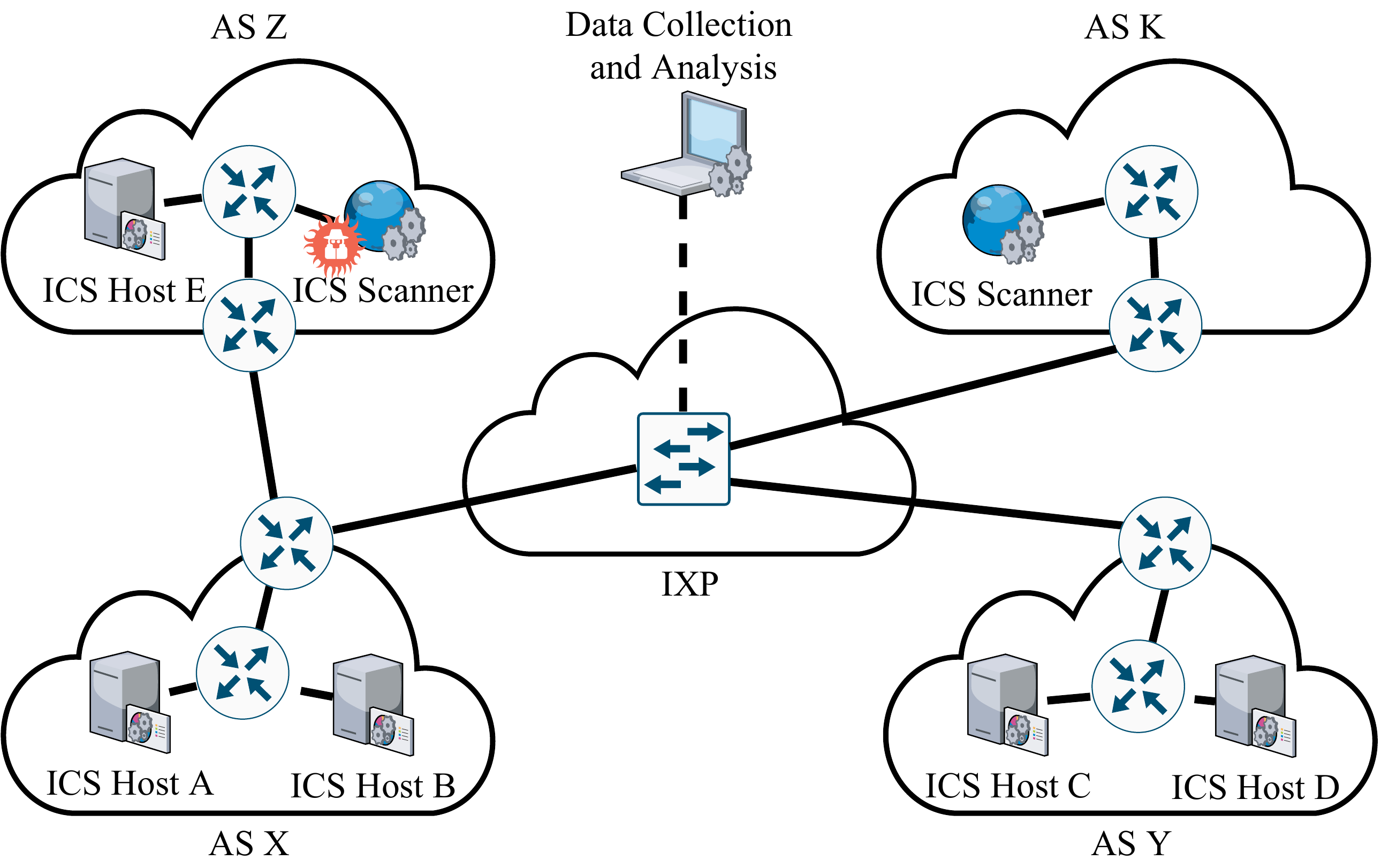}
\caption{System model: ICS at \textit{AS X}, \textit{AS Y}, and \textit{AS Z} communicate over the IXP. Scanners on the Internet (\textit{AS Z}, \textit{AS K}), which can be malicious or benign, look for exposed ICS services. The packets exchanged between two hosts belonging to \textit{AS Y} and \textit{AS Z} could still pass through the IXP, even if \textit{AS Z} is not directly connected to it. We sat within the IXP network where we are able to collect and analyze sampled packets.}
\label{fig:system_model}
\vspace{-1em}
\end{figure}

In our system model we assume that different ICSs are connected to a local AS (e.g., as customers of an ISP), and the ICSs are using industrial traffic over the Internet for supervision and control (depicted in Fig.~\ref{fig:system_model}).  
The IXP provides the capability to sample traffic, according to the sFlow standard~\cite{sflowstandard}, to understand the used protocols and involved parties. However, all the intra-AS communications and inter-AS communications that do not cross the IXP will not be identified by our collector.

In the considered model, an attacker could be a malicious end host located at another AS, or a Man-in-the-Middle attacker. The former attacker needs to be active (e.g., scanning the well-known ICS ports), while the latter can be passive or active. In particular, this last attacker should be able to sniff the traffic from the IXP, therefore it could be a malicious operator of the IXP, an insider who obtained the access at the IXP (e.g., the government), an ISP, or, more in general, an IXP partner.
The goals of the attackers are to learn process data (i.e., eavesdrop), and/or manipulate the ICS actively (e.g., by changing or injecting operational commands).

\subsection{ICS Packet Sampling}\label{subses:sampling_model}
We assume that the ICSs connected to the IXP are continuously sending traffic over the Internet, with a rate of at least one packet per minute. This is a reasonable assumption in line with other works~\cite{lemay2016providing, formby2016s}, since the ICSs continuously monitor processes with hard real-time constraints. In fact, these systems require frequent and constant polling-time communications to monitor the physical processes.

We also assume that the number of sampled packets $n$ is at least 10 times lower than the total number of packets $N$ passing through the IXP over the observation period $T$.
In this case, the sFlow sampling process can be modeled as a Binomial distribution $B(n,p)$~\cite{sflowstat,jedwab1992traffic} where $p$ is the probability of a successful event. We will discuss this assumption in Section~\ref{subsec:shodan_comparison}.

Under these assumptions, if an ICS is connected to the IXP and transmits 1 packet per minute, we can estimate the probability $\hat{p}$ that the sampled packet belongs to the ICS, as:

\begin{equation}\label{eqn:prob_ics}
    \hat{p}=\frac{1\cdot 60\cdot 24\cdot T}{N},
\end{equation}

where $T$ is the observation period, in days, and the numerator defines the overall number of packets sent from such ICS. In the limit case, we obtain $\hat{p}=1$, if $N=1\cdot60\cdot24\cdot T$. This corresponds to the case that we collected only packets of the ICS host under consideration.
The probability that we observe at least one packet with the aforementioned characteristics is:
\begin{equation}\label{eqn:1pkt}
    P(X\geq 1) = 1 - P(X=0).
\end{equation}

Generally, the probability to observe $k$ industrial packets is:

\begin{equation}\label{eqn:kpkts}
    P(X=k) =~{n\choose k}~\hat{p}^k~(1-\hat{p})^{n-k},
\end{equation}

where $n$ is the total number of sampled packets and $\hat{p}$ is defined in Eq.~\ref{eqn:prob_ics}.

\subsection{Research Questions, Challenges, and Goal}\label{subses:questions}
Our main goal is to investigate the practical use of ICS protocols over the Internet, in particular with respect to security. The main questions are: \emph{RQ1: How often are (insecure) ICS protocols used over the Internet? RQ2: How often are ICS services exposed to third parties, in addition to the intended use by legitimate parties? 
} 


The main challenge we faced is that third parties cannot directly observe legitimate use of ICS traffic unless they are routing the traffic (i.e., are in a Man-in-the-Middle position). Even in that case, efficiently filtering for ICS traffic out of large volumes of traffic can be challenging.

The outlined challenges raise the following additional research questions: \emph{If ICS protocols are used or exposed, can we identify such hosts using active scanning (e.g., \emph{Shodan}), or IXP/based traffic collection? To which degree are the results of both complementing each other?} In other words, \emph{RQ3: Is IXP active-scanning based enumeration of hosts a good estimator of (vulnerable) industrial traffic use on the Internet?}

For practical reasons, we focused our work on a geographical region in our country served by a specific IXP.

\subsection{Proposed Framework}\label{subses:pipeline}
\begin{figure}[t]
\centering
\includegraphics[width=\columnwidth]{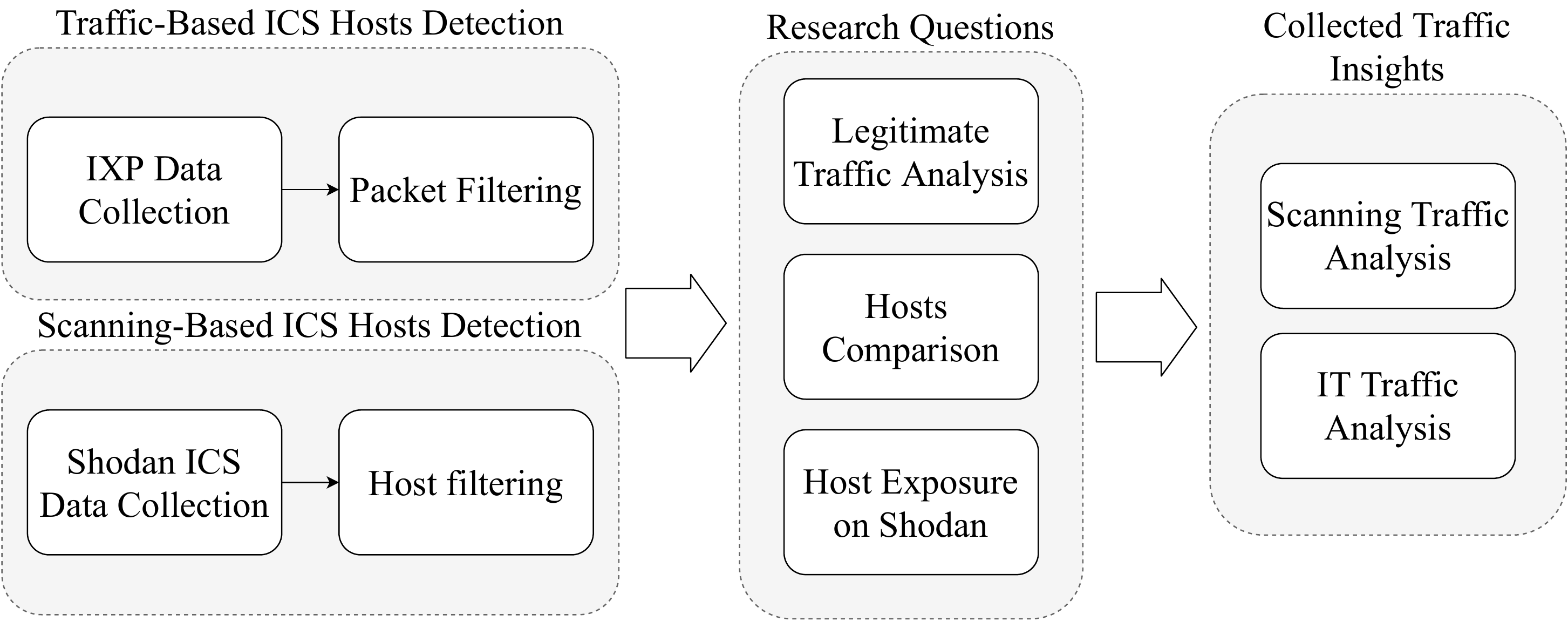}
\caption{The proposed framework to compare the data collected at the IXP and the information available on \emph{Shodan}.}
\label{fig:pipeline}
\vspace{-1em}
\end{figure}

To address the research questions proposed in Section~\ref{subses:questions}, we compared the properties of the industrial traffic passively collected at an IXP, with the information gathered by \emph{Shodan} which actively scan industrial ports.
We summarized the steps performed in Fig.~\ref{fig:pipeline}.
The first step consists in the data collection and processing.
However, the different nature of the two data sources (i.e., the IXP traffic and \emph{Shodan}) requires some consideration. In fact, while \emph{Shodan} monitors all the IP address space, the traffic captured by us is restricted to a geographical region, which makes the comparison unfair. To solve this, we defined a baseline to evaluate the two different approaches. The baseline is composed of the ICSs indexed on \emph{Shodan} which are the most likely to be observed from our positions, that are the ones who belong to ASs directly connected to the IXP.
Once the data were collected, we implemented a three-step filtering approach to extract the industrial traffic from the large dataset of network packets collected at the IXP. This procedure is based on well-known tools (e.g., Wireshark) and is able to identify both scanning and legitimate ICS activities. 

In the second step, we analyzed the results of the first step to answer the research questions. We analyzed the legitimate industrial traffic, compared the hosts identified with the baseline, and investigated the exposure on \emph{Shodan} of the hosts detected legitimately exchanging industrial traffic. Finally, to gather additional information about the current ICS security practices and threats, we investigated the data collected at the IXP to detect scanning activities and to deeply understand the network behavior and architecture (e.g., presence of a NAT) of the hosts. 
A port-scan approach cannot identify industrial services hidden behind a NAT, Firewall or VLAN. Instead, we can identify hidden industrial services by leveraging our IXP sampling based approach. The presence of both Industrial and non-industrial traffic from a single host can be an indicator of such hiding network mechanisms.

\section{Implementation}\label{sec:implementation}
In this section we present the implementation of our framework of analysis. In particular, Section~\ref{subsec:shodan_baseline} defines the baseline that we use to compare our approach to a scan-based one and Section~\ref{subsec:filtering} outlines the packet filtering approach used to identify the ICSs hosts.

\subsection{Collection of the \emph{Shodan} Baseline}\label{subsec:shodan_baseline}

We used \emph{Shodan} to identify the ICS hosts exposed to the Internet within the IXP area as baseline, due to popularity in related works~\cite{Al-Alami2018,Twente2019}, and because \emph{Shodan} supports significantly more ICS protocols with respect to other services such as Censys.
We define the IXP area as the set of ASs directly connected to the IXP itself. Due to the limited access to the \emph{Shodan} services we collected all the Italian ICS exposed according to the \textit{industrial-control-systems} category offered by the \emph{Shodan} platform. 
However, the list of protocols offered in such category is incomplete compared to the list of protocol dissector implemented in Wireshark and reported in Table~\ref{tab:wireshark_protocols}. In this table we reported the complete list of the industrial protocols of our interest, the communication ports used and if the current version of Wireshark (currently version is 3.0.5-1) is able to dissect them.
To address the limitations of the aforementioned category, for each of the missing protocol we designed a specific query based on the reference ports and common terms (e.g., \texttt{port:10001 country:"us" 
I20100} to discover ATG tank monitoring systems) collecting the resulting hosts. We performed the queries by leveraging Shodan API.
Finally, we selected the ICSs of our interest discarding the hosts that do not belong to an AS of the IXP area. We reported the results of the analysis in Table~\ref{tab:shodan_italy}.  

\begin{table}[t]
\centering
\footnotesize
\begin{tabular}{l|ccc}  \toprule
Protocol & \multicolumn{2}{c}{Port Ranges} & Wireshark \\
& TCP & UDP & \\ \midrule
AMQP  & 5671-5672 & - &  $\checkmark$ \\
ANSI C12.22  & 1153 & 1153 &  $\checkmark$  \\
ATG  & 10001 & - & \\
BACnet/IP  & - & 47808 &  $\checkmark$  \\ 
CoAP  & 5683 & 5683  & $\checkmark$  \\
Codesys  & 2455 & - & \\
Crimson v3   & 789 & - & \\
DNP3   & 20000 & 20000  & $\checkmark$  \\
EtherCAT  & 34980 & 34980  & $\checkmark$ \\
Ethernet/IP & 44818 & 2222  & $\checkmark$ \\
FL-net & - & 55000-55003  & \\
FF HSE & 1089-1091 & 1089-1091  & $\checkmark$ \\
GE-SRTP & 18245-18246 & -  \\
HART IP & 5094 & 5094  & $\checkmark$ \\
ICCP & 102 & - & $\checkmark$ \\
IEC60870-5-104 & 2404 & - & $\checkmark$ \\
IEC61850  & 102 & - & $\checkmark$\\
Modbus/TCP & 502 & - & $\checkmark$  \\
MELSEC-Q & 5007 & 5006  & \\
MQTT & 1883,8883 & - &  $\checkmark$  \\
Niagara Fox & 1911,4911 & - & \\
OMRON FINS & - & 9600  & $\checkmark$ \\
OPC UA & 4840 & - & $\checkmark$ \\ 
PCWorx  & 1962 & - & \\
ProConOS & 20547 & 20547  & \\
PROFINET & 34962-34964 & 34962-34964  & $\checkmark$ \\
S7comm & 102 & - & $\checkmark$ \\
Zigbee IP & 17754-17756 & 17754-17756  & $\checkmark$ \\
\bottomrule
\end{tabular}
\caption{Industrial protocols with relative ports, Wireshark dissector availability ( $\checkmark$ = wireshark is able to dissect it). }
\label{tab:wireshark_protocols}
\vspace{-1em}
\end{table}

\subsection{Packet Filtering}\label{subsec:filtering}

The large size of the traffic captured required the implementation of an automatic filtering approach. 
To identify the ICS protocols we applied a preliminary port-based filter, basing on the official documentation of the protocols and the ports list~\cite{ports} as main references. We reported the considered protocols in Table~\ref{tab:wireshark_protocols} followed by their port ranges and the Wireshark dissector availability. Then, due to the Wireshark dissector limitations, we applied the following three-step approach. 
\begin{table}[t]
\centering
\footnotesize
\begin{tabular}{lcccc}  
& \multicolumn{2}{c}{Italy} & \multicolumn{2}{c}{IXP Area} \\ \toprule
Protocol &   Hosts & \% &  Hosts & \%\\ \midrule
MQTT            & 1531 &  23.5  &   206 & 47.8 \\
Niagara Fox     & 1382 & 21.2 &   50 & 11.3 \\
Modbus/TCP       & 1346 & 20.6 &   70 & 15.9 \\
Ethernet/IP      & 456 & 7 &     16 & 3.6 \\
Siemens s7       & 417 & 6.4 &   37 & 8.4 \\
PCWORX           & 373 & 5.7 &      2 & 0.4 \\
CoAP             & 364 & 5.6 &       3 & 0.7 \\
Codesys          & 183 & 2.8 &    10 & 2.3 \\
BACnet/IP        & 170 & 2.6 &    26 & 5.7 \\
AMQP             & 140 & 2.1 &     4 &  0.9 \\
Omron FINS       & 83 & 1.3 &    6 & 1.4 \\
ATG              & 32 & 0.5  &   7 & 1.6  \\
OPC UA             & 13 & 0.2 &  1   & 0.2 \\
DNP3             & 11 & 0.2 &       & \\
IEC60870-5-104    & 6 &  0.09 &     &\\
ProConOS         & 5 & 0.07 &   &\\
GE-SRTP          & 5 & 0.07 &      1 & 0.2 \\
Crimson v3       & 2 & 0.03 &   & \\
MELSEC-Q         & 1 &  0.01    &      1 & 0.2 \\ \bottomrule
\end{tabular}
\caption{List of the ICS per-protocol hosts exposed in Italy and IXP area according to and the corresponding percentage over the total number.}
\label{tab:shodan_italy}
\vspace{-1em}
\end{table}

We reported the number of filtered packets at each step of the filtering process in Table~\ref{tab:filt_result}.
\begin{enumerate} 
\item Starting from the previously filtered packets, we identified the correctly dissected ICS packets by Wireshark, dividing scanning activities from legitimate activities.
    \begin{enumerate}
        \item We performed a deep packet inspection with Wireshark to identify all the Industrial Protocols packets supported and marked as not malformed;
        \item We cross-validated the resulting data using \textit{nDPI}~\cite{ndpi}, an open-source library able to dissect a wide range of industrial and non-industrial protocols, removing all the packets not tagged as non-industrial protocols;
        \item Then, we filtered out the packets where the IP-source is tagged as scanner by \textit{Greynoise}~\cite{greynoise}. \textit{Greynoise} is a security company that collects, labels, and analyzes Internet-wide scan and attack data, and provides the access to such information to users via API\footnote{The classification methodology of Greynoise in not public and this could affect the deduced results and reproducibility of the experiment.}.
        \item We considered the difference between the results of b) and c) as \textit{legitimate ICS traffic};
    \end{enumerate}
\item In the second step we further processed the initial port-based filtered packets. The goal is to identify additional scanning activities targeting the ports of the industrial protocols not supported by Wireshark.
    \begin{enumerate}
        \item We used Wireshark to remove all the not malformed, industrial and non-industrial protocols (e.g., SSL, HTTP), keeping the packets generally tagged as TCP or UDP (e.g., SYN packets scanning industrial ports);
        \item We used \textit{nDPI} to remove additional non-industrial traffic from the previous results;
        \item Then we used \textit{Greynoise} to identify scanning activities;
    \end{enumerate}
\item In the third step we gather the results of first two steps. 
    \begin{enumerate}
        \item We merged the scanners activities identified in 1.c and 2.c, and we tagged them as \textit{ICS scanners};
        \item We considered the sum of the legitimate ICS traffic (i.e., 1.d) and the ICS scanners (i.e., 3.a) as \textit{ICS traffic}.
    \end{enumerate}
\end{enumerate}



\begin{table*}[!t]
\footnotesize
\resizebox{\textwidth}{!}{%
\begin{tabular}{l  @{\hskip 8pt} | llllllllllllll llllllllllll|l }

\multicolumn{1}{l}{} &	\rot{MQTT}	&	\rot{Zigbee IP}	&	\rot{EtherCAT}	& \rot{PROFINET} & \rot{IEC60870-5-104} & \rot{ANSI C12.22} & \rot{Ethernet/IP} & \rot{Modbus/TCP} & \rot{AMQP} & \rot{OPC UA} & \rot{CoAP} & \rot{FF HSE} & \rot{ATG*} & \rot{BACnet/IP} & \rot{Codesys\*} & \rot{GE-SRTP*} & \rot{ICCP/IEC61850/s7} & \rot{MELSEC-Q*} & \rot{Niagara Fox\*} & \rot{OMRON FINS} & \rot{PCWorx*} & \rot{HART IP} & \rot{Crimson v3*} & \rot{DNP3} & \rot{FL-net*} & \rot{ProConOS*} & \rot{Overall}\\
\midrule 
$h$ & 62 & 32 & 16 & 15 & 10 & 10 & 8 & 8 & 6 & 4 & 3 & 2 & 0 & 0 & 0 & 0 & 0 & 0 & 0 & 0 & 0 & 0 & 0 & 0 & 0 & 0 & 176\\
$h_S$ & 206 & 0 & 0 & 0 & 0 & 0 & 16 & 70 & 4 & 1 & 3 & 0 & 7 & 26 & 10 & 1 & 37 & 1 & 50 & 6 & 2 & 0 & 0 & 0 & 0 & 0 & 442\\
$i$ & 1 & 0 & 0 & 0 & 0 & 0 & 0 & 1 & 0 & 0 & 0 & 0 & 0 & 0 & 0 & 0 & 0 & 0 & 0 & 0 & 0 & 0 & 0 & 0 & 0 & 0 & 2\\
\midrule
\end{tabular}}

\caption{Comparison between the number of ICS hosts found with our approach ($h$) and by \emph{Shodan} ($h_{S}$) within the IXP area. The value $i$ represents the number of hosts common to the two approaches. * = Wireshark dissector not available.}
\label{tab:results_protocols}
\vspace{-1em}
\end{table*}

\section{Results}\label{sec:results}

We collected data for a period of 31 days from the 14th of January 2020 to the 14th of February 2020 at the considered IXP. According to the sFlow standard, the traffic was sampled with a sampling rate of $2^{-12}$ and packet truncation at 128 bytes. The sampling process provides an estimation of the effective traffic passing through the exchange point~\cite{sflowstat}, while the truncation gives access to the full link layer, network layer, transport layer, and few bytes of the payload. The collection resulted in $\sim$1.6B packets for more than 189GB of data. All the source and destination IPs were anonymized on-the-fly during the capture process. Furthermore, for privacy concerns, we excluded the entire payload from the analysis.

In the following section, we present the analysis results of the traffic capture obtained after the collection phase. In particular, Section~\ref{subsec:legitimate} discusses the legitimate ICS traffic, Section~\ref{subsec:shodan_comparison} compares our approach with \emph{Shodan}, while Section~\ref{subsec:shodan_results} reports an analysis of the \emph{Shodan}-indexed hosts.

\begin{table}[t]
\centering
\footnotesize
\begin{tabular}{l|cc}  \toprule
 & Packets & \% \\ \midrule
Total & $\sim$1.6B &  \\
After Port-based filtering & 43584 & 0.0027 \\\hline 
\textbf{Step 1} & & \\
a) Wireshark filtering & 3188 & - \\
b) nDPI validation & 2075 & 65.1 \\
c) Scanning activities & 1360 & 42.6 \\
d) Legitimate ICS traffic & 715 & 22.4 \\ \hline
\textbf{Step 2} & & \\
a) Wireshark filtering & 32741& -\\
b) nDPI validation & 26171 & 80\\
c) Scanning activities & 3019 & 9.2\\\hline
\textbf{Step 3} & & \\
a) Total ICS scanners    & 4379 & -\\
b) Total ICS traffic  & 5094 & - \\ \hline \midrule 
 & \multicolumn{2}{c}{Overall industrial traffic \%} \\ 
& \multicolumn{2}{c}{w.r.t. total one} \\ \midrule

ICS traffic & \multicolumn{2}{c}{$\sim$0.0003}  \\
ICS scanners & \multicolumn{2}{c}{$\sim$0.0003} \\
Legitimate ICS traffic & \multicolumn{2}{c}{$\sim$0.00005}\\ \bottomrule
\end{tabular}
\caption{Number of packets filtered step by step. The percentage represents the number of packets left compared o the previous step (i.e., the previous row).}
\label{tab:filt_result}
\vspace{-1em}
\end{table}

\subsection{Legitimate ICS Traffic}\label{subsec:legitimate}

We identified 168 different ICS endpoints and 12 different industrial protocols, with 8 hosts using two different ICS protocols. The percentage of legitimate ICS traffic identified over the total number of packets is $5\cdot10^{-5}$.
Overall, MQTT and AMQP protocols count more than 55\% of the legitimate ICS traffic. This result is not surprising since both the protocols are widely used for IoT communication in non-industrial environments. ANSI C12.22 covers almost 21\% of the total number of packets, followed by Modbus/TCP with 9\% and Zigbee with 8\%. Other common ICS protocols are Profinet, Ethercat, IEC60870-5-104 and Ethernet/IP together count about the 7\%. Another interesting result is that despite MQTT official documentation~\cite{mqtt_oasis} specifies port 8883 and AMQP port 5671 respectively for communicating over TLS, our results show that all the MQTT and AMQP communication rely on insecure ports, leading to known vulnerabilities~\cite{Andy2017, Panchal2019}.

\subsection{Comparison with the \emph{Shodan} Baseline}\label{subsec:shodan_comparison}
To address \emph{RQ3} we compared the hosts identified in Section~\ref{subsec:legitimate} with the hosts identified on Section~\ref{subsec:shodan_baseline}. We define $H$ as the set of ICS hosts detected by us and $H_S$ as the set of ICS hosts identified by \emph{Shodan}. 
We also compute the number $i = |H \cap H_S |$ of ICS hosts detected by both two approaches. 






We reported the results in Table~\ref{tab:results_protocols}. Among the 176 hosts $h$ identified in the Table, 12 hosts are duplicated (i.e., the same use more than one protocol), while among the 402 $h_S$, two hosts are duplicated.
Despite we detected an overall amount of 168 unique ICS hosts compared to 440 by \emph{Shodan}, only 2 hosts were common to both the approaches, respectively an MQTT and a Modbus/TCP endpoint, meaning that the two methods are complementary.
Our approach 
detected more hosts than \emph{Shodan} for 8 protocol. The \emph{Shodan} port-scanning approach, instead, detected more hosts for 12 protocols. 

We further investigated why we detected only two endpoints in common with \emph{Shodan}. To do so, we extracted from the dataset collected at the IXP all the packets that come from, or are directed to, an element of $H_S$. Results show that the 16.3\% of $|H_S|$ (72 hosts out of 440) were captured during our analysis but were not leveraging industrial communication. This means that they correspond to false positives, since they were not actually using industrial protocols, or that their packet rate transmission was below our threshold of 1 packet per minute (defined in Section~\ref{subses:sampling_model}).

We relied on sampling rate the assumptions of Section~\ref{subses:sampling_model}, which are verified since $n<<10N$ (i.e., the sampling rate is $2^{-12}$ and $N\approx 2^{12}n$), to compute the probability that we miss all the packets of an ICS host (\ref{eqn:PX0}) and the probability that we observe at least one packet of an ICS host (\ref{eqn:PXg1}) in the sampling period. 
In particular, given $T=31$ days, $n=1599431398$ sampled packets:
\begin{equation}\label{eqn:PX0}
P(X=0)=1.848 \cdot 10^{-5},
\end{equation}
\begin{equation}\label{eqn:PXg1}
P(X\geq 1)= 1 - P(X=0) = 0.999.
\end{equation}
According to the previous results, the probability we missed an ICS host that respect our assumptions is negligible. A possible explanation is that the 72 hosts indexed as ICS by Shodan are false positive (e.g., they are general devices with exposed ICS ports) or ICS hosts but are not currently active.

\subsection{Hosts Exposure on \emph{Shodan} }\label{subsec:shodan_results}

Among the ICS hosts involved in legitimate ICS communication, we are interested in counting how many of them were also indexed by \emph{Shodan}, and what kind of information \emph{Shodan} was able to gather (i.e., \emph{RQ2}).
We found that 64.3\% of such hosts were successfully identified by \emph{Shodan} and 11\% of them were found with ICS ports exposed, more specifically Modbus/TCP, MQTT, and AMQP. 

In Table~\ref{tab:shodan_product_port} we reported the Top-5 exposed services and the Top-5 exposed ports based on the \emph{Shodan} collected data.
Due to the high percentage of Web Servers detected, we investigated deeply to understand what kind of services these devices were exposing. We found IP Cameras, Printers, Routers, and Network Attached Storage (NAS) login pages other than energy monitoring and alarm systems.
Note that in this work, due to the potential criticalities of the involved systems, we were not interested in performing any active penetration test to the identified devices for security and privacy concerns. For this reason, we analyzed the Common Vulnerabilities and Exposures (CVEs) information provided by \emph{Shodan} to identify possible vulnerabilities caused by unpatched systems or well-known critical services.
We found that 10.2\% of the systems indexed were affected by at least one CVE. We then associated with each vulnerability the corresponding Common Vulnerability Scoring System (CVSS) v2.0 score~\cite{cvss}. According to the National Institute of Standard and Technology (NIST) the severity of a score between \textit{0.0}-\textit{3.9} is considered low, \textit{4.0}-\textit{6.9} medium and \textit{7.0}-\textit{10.0} high.
We found an overall amount of 207 CVEs, 30\% of which have a score greater than 7.0 affecting 81\% of the vulnerable hosts, 4.3\% greater than 8.0, 2.9\% greater than 9, and 2.4\% equal to 10.0 all affecting 27.3\% of the vulnerable hosts. We must note that all these vulnerabilities can be exploited remotely.

\begin{table}[t]
\centering
\footnotesize
\begin{tabular}{lc||lc}  
Product & \% & Port & \% \\ \toprule
1) MikroTik bandwidth-test server   &  25 & 1) 443   &   11.8 \\
2) Apache httpd                     &  12.5 & 2) 80  &   11.5 \\
3) nginx                            &  9 & 3) 2000 &     7 \\
4) Open SSH                         &   9 &  4) 8080   &    6 \\
5) MQTT                             &   7.9 & 5) 22  &   4.1\\
\bottomrule
\end{tabular}
\caption{Top-5 ports and services detected by \textit{Shodan} and  percentag eof hosts exposing them.}
\label{tab:shodan_product_port}
\vspace{-1em}
\end{table}

 \subsection{Summary of Main Results}
\noindent\textbf{RQ1:} \emph{How often (insecure) ICS protocols are used over the Internet?} We observed 12 different industrial protocols during our collection period. By analyzing these 12 protocols, 6 of them (i.e., EtherCAT, PROFINET, IEC60870-5-104, Ethernet/IP, Modbus/TCP, FF HSE) do not implement any encryption, authentication or integrity protection features by design and were used by 59 hosts. In addition, protocols such as MQTT and AMQP support TLS (enabling confidentiality and authentication), but this was not implemented in practice. The use of insecure protocols and missing use of TLS affected an overall amount of 127 hosts, meaning that 75.6\% of the hosts are using vulnerable ICS communications.

\noindent\textbf{RQ2:} \emph{How often are ICS services exposed to third parties, in addition to the intended use by legitimate parties?} We found only a small subset of hosts that we identified as legitimately using ICS protocols (i.e., 7.1\% corresponding to 12 hosts) also have ICS protocols ports exposed to the public Internet. Furthermore, for the hosts that we identified as legitimately using ICS protocols, we found that a good subset (i.e., 64.3\% corresponding to 108 hosts) also has general IT ports exposed to the Internet. By analyzing these 108 hosts, 11 of them were affected by at least a CVE, instead 9 of them were affected by different CVEs with a CVSS score greater than $7.0$.

\noindent\textbf{RQ3:} \emph{Is IXP active-scanning based enumeration of hosts a good estimator of (vulnerable) industrial traffic use on the Internet?} Table~\ref{tab:results_protocols} shows that \emph{Shodan} finds three times as many hosts as our method, while the $i$ value we calculated indicates that only about 1.2\% of the hosts collected by our framework were also detected by \emph{Shodan}. This means that \emph{Shodan} missed most hosts that are actually implementing ICS protocols. 
We can conclude that while \emph{Shodan} returns a higher number of hosts, it is largely unable to identify hosts that actually use industrial protocols for legitimate applications (identified instead by the traffic analysis) even with a forced manual direction to the right hosts. This may be due to the presence of NATs or other mitigation mechanisms which block the direct scan on the host port.

\section{Additional Analysis and Insights}\label{sec:extra_res}

In the following section we reported further results we obtained beyond the research questions defined in Section~\ref{subses:questions}. 
In particular, Section~\ref{subsec:scanners} provides a detailed analysis of the origin of scanning campaigns. Section~\ref{sec:mashup} reports an analysis of the hosts implementing both Industrial communication and non-Industrial communication. Finally, Section~\ref{subsec:validation} presents a validation procedure for the IXP on which we relied on.

\subsection{Scanning Activities}\label{subsec:scanners}

In this analysis, we associated any scanning activity identified during the packet filtering process to the relative ICS protocol, according to the targeted port. 

\begin{figure}[t]
\centering
\includegraphics[width=\columnwidth, trim={0.85cm 0 1.3cm 1.8cm}, clip]{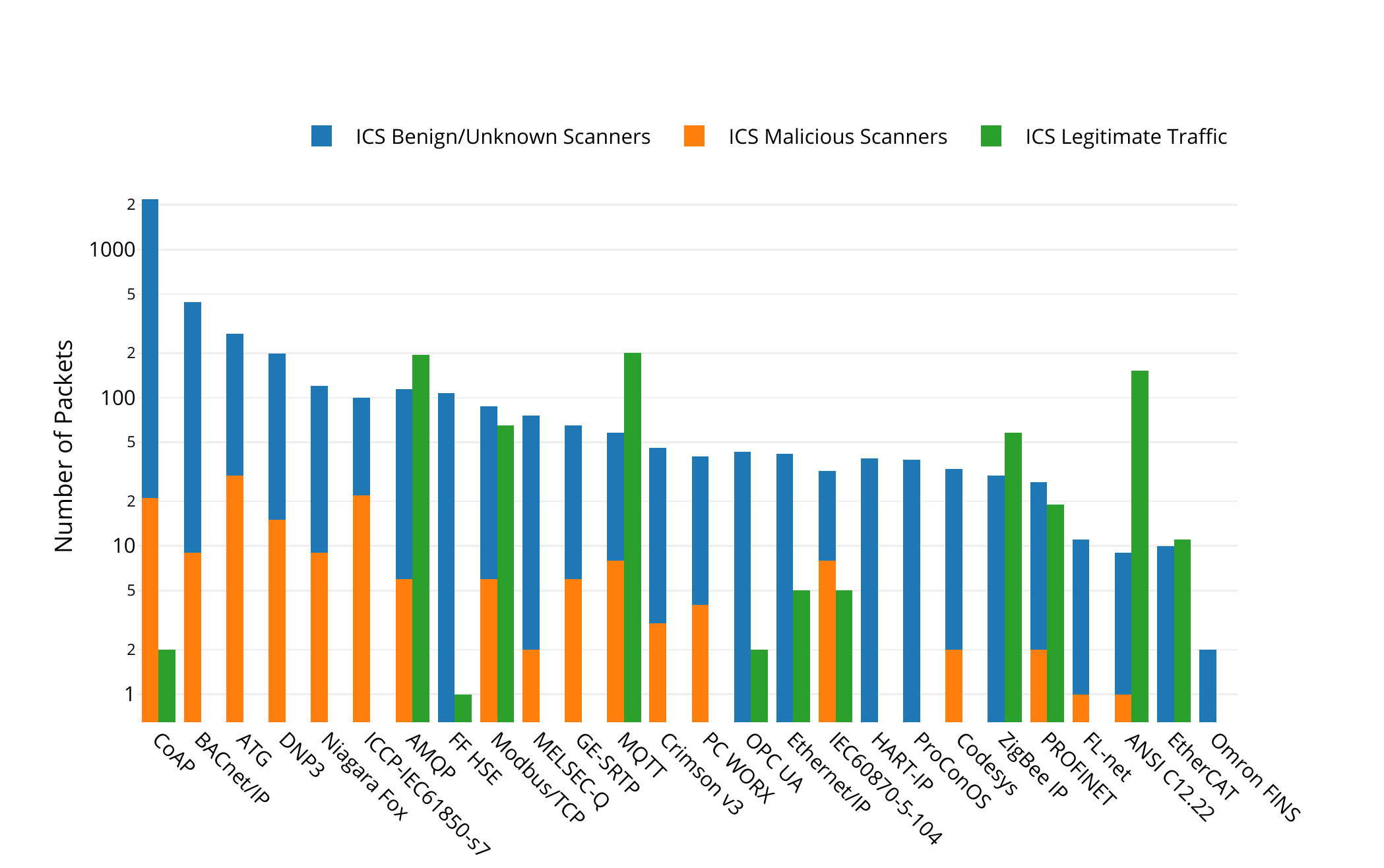}
\caption{Protocols found among the legitimate and scanning ICS traffic. The Y-axis is log-scaled.}
\label{fig:charts}
\vspace{-1em}
\end{figure}

\parag{Scanned protocols} In Fig.~\ref{fig:charts} we reported an overview of the protocols scanned by malicious actors. 
We identified a total of 442 different IPs performing scanning activities. 
The most scanned port was 5683 used by CoAP with more than 50\% of the scan packets, followed by BACnet/IP, Automated Tank Gauge (ATG) systems and DNP3. 
Almost 30\% of the scan packets were designed with protocol-specific requests, like \textit{readProperty} function for BACnet/IP, \textit{GET /.well-known/core} for CoAP, \textit{List Identity} for Ethernet/IP, \textit{Session Initiate Request} for HART IP and \textit{Controller Data Read} for OMRON FINS. The remaining 70\% of the packets consisted of 61\% of UDP packets, 32\% of simple SYN packets, 35\% of RST and less of the 1\% for SYN-ACK and other combination, which confirms an established TCP connection. 

\parag{Scanning actors} We used the \textit{Greynoise} platform~\cite{greynoise} to tag the scanners as Malicious, Benign or Unknown, according to their logged activities.
\textit{Greynoise} reported the $3.5\%$ of the host as Malicious, $37.4\%$ as Benign and the remaining $59\%$ as Unknown. 
For instance, a malicious actor was associated with a behavior indicating a Mirai or a Mirai-like variant infection, while another one was found opportunistically scanning for Siemens PLC devices. 
We also identified well-known services that periodically scan the Internet address space such as \emph{Shodan} and \textit{Censys}~\cite{censys}. 
In particular the Top-5 most frequent Scanner actors we identified are composed by the 35\% by Censys, the 4.3\% by \textit{Stretchoid}, the 4.3\% by \emph{Shodan}, the 4.3\% by \textit{Net Systems Research}, and the 3.9\% by \textit{BinaryEdge}, while the 44.1\% are unknown actors. 



\begin{figure}[t]
\centering
\footnotesize
\includegraphics[width=\columnwidth, trim={2.1cm 2.6cm 0.6cm 3.0cm}, clip]{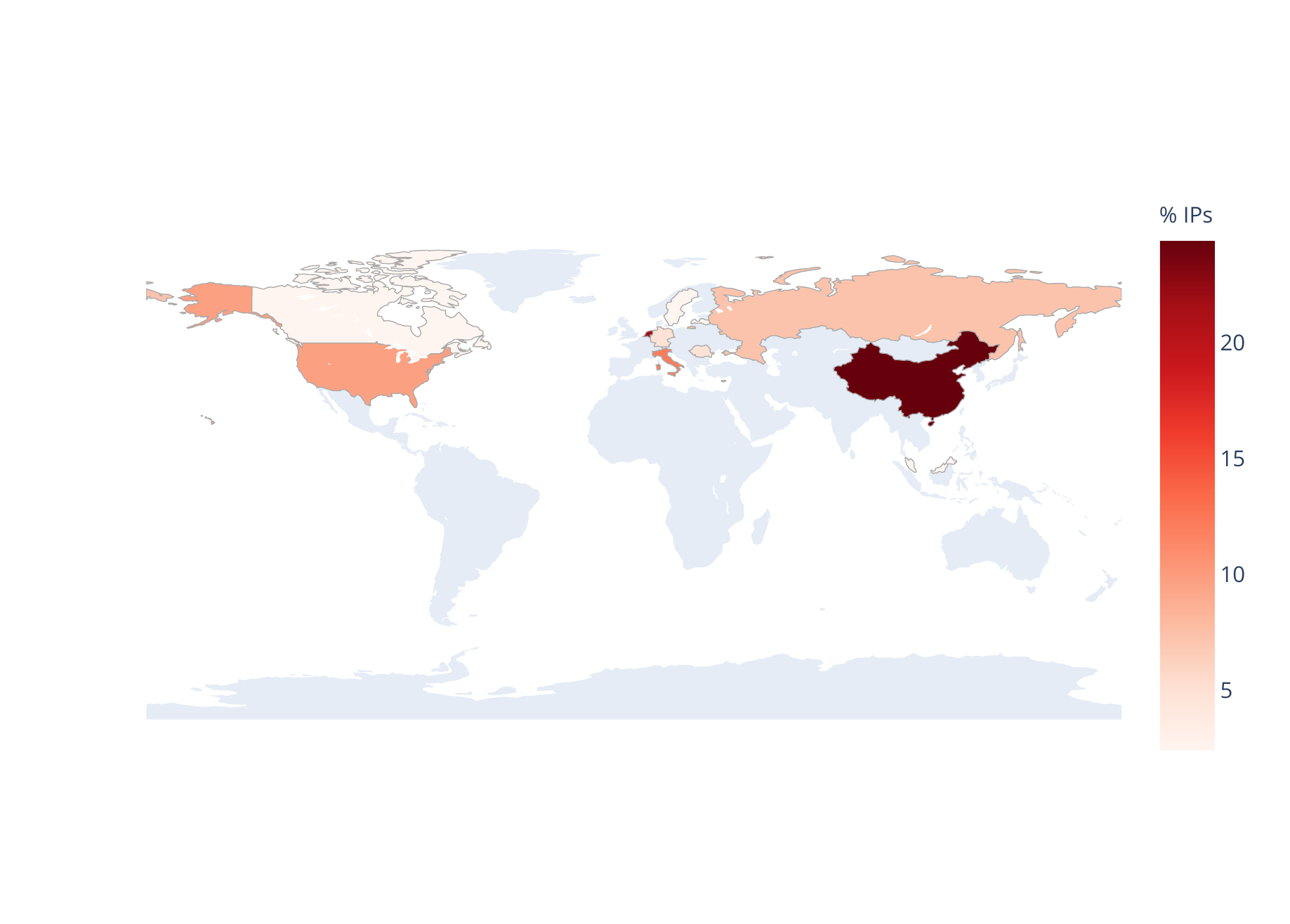}
\caption{Heatmap of the malicious scanning activities origin.}
\label{fig:malicious_scanners}
\vspace{-1em}
\end{figure}

\parag{Malicious scanners} We observed that the ATG port is the most scanned by malicious actors, even more than well-known ICS protocols such as BACnet/IP, Modbus/TCP, and DNP3. 
Remote access to the control port of an ATG could provide an attacker the ability to reconfigure alarm thresholds, reset the system, and to disrupt the operation of the fuel tank~\cite{automated_tank_gauge,wilhoit2015little}. However, since ATG is mainly used in the USA, this amount of scan traffic can be due to port 10001 shared with other services. According to~\cite{port10001}, several malware leverages this port to spread over the devices, furthermore, \emph{Shodan} reports that almost all the devices with such exposed port are network antennas. 
We reported in Fig.~\ref{fig:malicious_scanners} a heatmap with the origin of the IP associated with each scanning packet. The scanning activities come from 30 different countries. The 24.4\% of the malicious actors come from China, 22\% Netherlands, 12\% Italy, 10\% USA, and 7\% Russia, while 8 other countries count for less than 5\%.
However, we must note that the authors of the scanning campaigns could also hide their source by using, for instance, a VPN. In this case, the IP origin represented in the map corresponds to the last VPN hop.

\begin{table*}[t]
\centering
\footnotesize
\begin{tabular}{l|lclc|lclc}  
 & \multicolumn{4}{c}{Packet-based} & \multicolumn{4}{c}{Flow-based} \\ \toprule 
& \multicolumn{2}{c}{Overall} & \multicolumn{2}{c}{ICS-to-ICS} & \multicolumn{2}{c}{Overall} & \multicolumn{2}{c}{ICS-to-ICS} \\ \toprule
\multirow{8}{0.13\textwidth}{Before IANA mapping} 
    & Protocol & \% & Protocol & \% & Protocol & \% & Protocol & \% \\ \midrule
    & TLS        &       50.7 & UDP     &   47.5     & TLS &     45 & TCP & 17.8 \\
    & HTTP       &       41.3 & TCP     &     40.1    & HTTP &     23.7 & TLS  &   15.8 \\ 
    & UDP        &        4.8 & OpenVPN    &   8.8     & TCP &     18.4 & UDP & 8.2  \\
    & TCP       &         2.7 & TLS     &   1.8       & UDP &     6.7 & ICMP & 4.1 \\
    & DNS        &        0.1 & SIP  &   1.2     & ICMP & 3 & RTCP & 3.1 \\ \midrule
\multirow{5}{0.13\textwidth}{After IANA mapping} 
   & TLS        &        50.7 & TCP     &    36.4     & TLS &     42.9 & TLS & 1 \\
    & HTTP       &       41.3 & UDP     &     16        & HTTP &     22.6 & TCP  &   0.6 \\ 
    & STUN       &        3.5 & OpenVPN     &   8.8     & XMPP &     6.6 & UDP & 0.4 \\
    & XMPP        &       1.7 & RSF-1 clustering  & 5.4 & HP V.ROOM &  5.5 & Reserved & 0.4  \\
    & UDP        &        0.8 & ActiveSync  &   2.4     & TCP & 3.2  & ICMP & 0.3 \\ \bottomrule
\end{tabular}
\caption{Top-5 non-industrial protocols.}
\label{tab:overall_mixitot_traffic}
\vspace{-1em}
\end{table*}

\subsection{IT Traffic}\label{sec:mashup}

We identified that more than 91.6\% of the industrial endpoints were also communicating via non-industrial protocols. 
This can be due to the use of more than one protocol by a single device, the presence of exposed IT services in ICS devices, or by the presence of a NAT on the network border that manages the traffic incoming or outgoing from the enterprise and manufacturing zone. In the first column of Table~\ref{tab:overall_mixitot_traffic} we reported that almost half of the traffic consists of encrypted TLS traffic, which could be due to the use of HTTP over TLS or to other secure communications such as VPNs. Moreover, HTTP covers 41.3\% of the overall traffic, DNS covers 0.1\%, while other interesting findings not mentioned in the table that represents less than 1\% are OpenVPN, ESP, Wireguard, STUN, BitTorrent, FTP, and Telnet. 

Due to the high amount of non-industrial traffic, we investigated if such behavior happened also between two endpoints of a legitimate ICS communication. 
We found out that 69.5\% of the legitimate ICS endpoints exchange also non-industrial traffic.  
In order to have a clear view of the identified protocols, we applied a flow-based approach, since the high amount of packets sent from a single host could affect significantly the statistics.
As we can see in Table~\ref{tab:overall_mixitot_traffic}, this behavior is strongly evident in the amount of HTTP traffic, where just 23.7\% of the hosts were using the HTTP protocol with respect to 41.3\% found on the packet-based approach. Moreover, to reduce the number of the not precisely tagged TCP protocol, we mapped the lower port of each communication with the corresponding port registered by the Internet Assigned Numbers Authority (IANA). This approach significantly changed the results of the flow-based approach for the two ICS-endpoints analysis. We must also note that the low percentages obtained are due to the high frequency with which some ports changed within the same communication.

\subsection{Validation}\label{subsec:validation}
To verify the correct functioning of our environment, we injected self-crafted traffic into the IXP. 
To do this, we deployed a Modbus/TCP server and a \textit{Mosquitto} MQTT broker in an Amazon EC2 server instance. Instead, within the IXP network we deployed a synchronous Modbus/TCP and a MQTT client based on \textit{pymodbus} and \textit{paho-mqtt} python modules. During the communication between clients and servers, the Modbus client sends a \textit{Write Single Register} request and the MQTT client sends a \textit{Publish Message} request. Considering that the 3-way handshakes were already accomplished, the overall amount of packets for each transaction is the following: 
\begin{itemize}[noitemsep,topsep=0pt,parsep=0pt,partopsep=0pt]
    \item Modbus/TCP:
            \begin{enumerate}[noitemsep,topsep=0pt,parsep=0pt,partopsep=0pt]
            \item Client sends \textit{Write Single Register} request;
            \item Server sends \textit{Write Single Register} response;
            \item Client sends \textit{TCP ACK} to server;
            \end{enumerate}
    \item MQTT:
            \begin{enumerate}[noitemsep,topsep=0pt,parsep=0pt,partopsep=0pt]
            \item Client sends \textit{Publish Message} request;
            \item Server sends \textit{TCP ACK} to client;
            \end{enumerate}
\end{itemize}

We sent 10 \textit{Write Single Register} and 10 \textit{Publish Message} requests per second for 24 hours. In this topology, the sFlow Agent samples just the traffic outgoing from the IXP network. It resulted in 2572852 packets exchanged, 1477915 outgoings and 346 of which were successfully sampled: 164  MQTT requests, 91 Modbus/TCP requests and 91 Modbus TCP ACK. Furthermore, all the packets were correctly dissected by Wireshark. The sampling rate computed as the ratio between the sampled packets and the overall outgoing traffic results in about $2.3411\cdot10^{-4}$ packets, which is what we expected, considering an sFlow sampling rate of $2^{-12}$. 
This validation confirms the correct functioning of our IXP environment, the correct sampling functioning of sFlow and the correct dissection function of the network packets by Wireshark.
\section{Discussion}\label{sec:discussion} 

\parag{Comparison of Approaches} Our traffic analysis approach give us a very different point of view with respect to \emph{Shodan}. While \emph{Shodan} collects data performing active port scanning and fingerprinting of the exposed services, having thus a wider overview of the exposed hosts, our approach allows identifying hosts currently active and communicating. 
Our analysis shows that \emph{Shodan} identified just 2 hosts actually exchanging industrial traffic. 
There are three possible explanations for this result: the first one is that the systems were hidden behind a firewall, NAT, VLAN or were leveraging other port scanning mitigation techniques (e.g., filtering IP of scanning services), the second one is that the hosts indexed on \emph{Shodan} were inactive during our measurement time window, and the third is that the packet transmission rate was low enough to avoid sampling detection (i.e., less than 1 packet per minute).

\parag{ICS Communications Security} The 98.8\% of the ICS that we were able to identify were not recognized as ICS by \emph{Shodan}. 
The 50\% of the ICS protocols identified in the legitimate traffic was not implementing any encryption mechanism due to insecure protocols by design (e.g., Modbus/TCP) or bad configurations (i.e., MQTT, AMQP), exposing the whole systems to potential attacks. 
The 91.6\% of hosts that exchange both industrial traffic and non-industrial traffic could be exploited by an attacker to investigate which protocols are used and what is the payload of the packets. For instance, the attacker could gather information about the servers that are commonly involved, analyze the DNS, FTP, and HTTP content to collect information about the employees, and use social engineering techniques to spoof them.

\parag{Limitations} Our approach presents some limitations. Therefore, as explained in Section~\ref{subses:sampling_model}, it is possible that sFlow was not able to detect some active ICS host since the accuracy of sFlow is dependent upon the sampling rate. The sampling rate also does not allow us to perform any encrypted traffic analysis since we could not record the entire communication flow. Moreover, the sFlow protocol truncates the packets at 128 bytes. This last limit, together with the fact that the Wireshark dissector does not support some protocols of interest, may lead to an incomplete analysis of the traffic. In our environment, the possible false negatives and false positives obtained are due to the third-party tool used in the extraction process (i.e., Wireshark, nDPI, and Greynoise), and this makes it impossible to verify the ground-truth. Finally, if a host implements a dynamic port range, instead of using the standard ports reported in Table~\ref{tab:wireshark_protocols}, both our and \emph{Shodan} approaches fail to correctly identify the host because both the approaches are port-based, leading to additional false negatives. More precisely, our approach relies on port-based inspection in two phases: i) during the preliminary port-based filtration; ii) during the Wireshark deep-packet inspection (i.e., Wireshark fails in identifying protocols that use non-standard ports).

\section{Related Work}\label{sec:related}

\parag{ICS Security}The security of ICS is a widely studied research topic. Many research groups study new security solutions to implement in order to mitigate the threats to which ICSs are exposed to. The Anomaly detection systems represent a low cost and effective solution. Anomaly detection systems do not require any hardware substitution from the point of view of the company. Indeed they aim to monitor the state of the system passively, focusing, for example, on the physical state of the network~\cite{evaluationml}, network traffic~\cite{anton2018evaluation} or considering both of them~\cite{kingfisher}, and raising an alert when the normal behavior of the system is violated.

\parag{Scan-Based Analysis}If, on one hand, there are many contributions in literature which present security solutions for the ICSs, on the other hand, there are not many contributions that analyze their current security implementation.
Several works~\cite{Al-Alami2018,Twente2019} leveraged public available information, like \emph{Shodan}, to identify internet-reachable industrial devices, while~\cite{Mirian2016} and~\cite{Feng2016} manually performed an active scan of the IPv4 address space. For a better understanding of the threat landscape, Serbanescu et al.~\cite{Serbanescu2015} deployed a low-interaction honeynet, also showing how the number of attacks increased after being indexed by \emph{Shodan}. 

\parag{Industrial Traffic Analysis} There is not much literature on industrial traffic analysis. In~\cite{Nawrocki2019} the authors, investigated the industrial communications passing through an IXP and an Internet Service Provider (ISP). During their analysis, they were able, via correlation techniques, to identify scanning activities, possible firewall implementation other than unprotected industrial traffic. 
In our work, we are not interested in implementing any new tool for identifying industrial traffic, however we wanted to exploit a wider range of protocols, so we proposed a slightly different packet filtering process to identify possible Industrial communications and more scanning activities. From a security point of view, we analyzed the critical issues deriving from implementing both industrial and non-industrial communication within the same network infrastructure and we argued what are the \emph{Shodan} and \emph{Shodan}-like services limits compared to a wide-view analysis of sampled traffic. 


\section{Conclusions}\label{sec:conclusion}


The increasing using of insecure industrial protocols through the Internet exposed ICS and critical infrastructure to a wide range of cyber threats. Active scanning of the IP address space performed by services such as \emph{Shodan} is a common practice to detect exposed ICS, however it does not properly represent the real use of insecure industrial protocols. In this paper, we addressed three research questions to investigate the current state of the art use of such protocols over the Internet by industrial systems. To do this, we proposed, implemented and validated an analytic framework to detect legitimate industrial traffic communication and scanning activities, based on a 31 days long sampled traffic capture collected at a local IXP. We compared our results with the information available on \emph{Shodan}, proving that Shodan is not enough. In fact, while \emph{Shodan} was able to identify a higher number of hosts, it detected the only 1.2\% of the hosts found by us. We also show that 64.3\% of the hosts have IT services exposed, 11 of which have an alarming CVE vulnerability score, and that 75.6\% of the industrial protocols implemented to communicate over the Internet do not implement any security feature. 

Additional analysis, such as the analysis of the IT traffic, confirmed the convergence of the Information Technology and Operational Technology networks in many systems (in our case the 91.6\% of the identified hosts), providing a deeper point of view of the network architecture and security, such as the high rate of unencrypted IT traffic and insecure protocols. These network vulnerabilities could be exploited by malicious users, who constantly perform scanning campaigns, as our results show.
Finally, we validated our system model by auto-injecting our traffic showing that it respects our assumptions.


\bibliographystyle{IEEEtran}
\bibliography{samplepaper}
\end{document}